\documentclass{article}
\usepackage{amsmath,amsfonts}
\usepackage{algorithmic}
\usepackage{array}
\usepackage[caption=false,font=normalsize,labelfont=sf,textfont=sf]{subfig}
\usepackage{textcomp}
\usepackage{stfloats}
\usepackage{url}
\usepackage{verbatim}
\usepackage[sorting=none]{biblatex}
\usepackage[dvipsnames]{xcolor}
\usepackage{graphicx}
\usepackage{enumitem}
\def\BibTeX{{\rm B\kern-.05em{\sc i\kern-.025em b}\kern-.08em
    T\kern-.1667em\lower.7ex\hbox{E}\kern-.125emX}}
\usepackage{balance}
\usepackage[normalem]{ulem}
\usepackage{hyperref}
\hypersetup{
	allcolors=blue!40!black,
    colorlinks=true,
    pdftitle={SourceDataSelectionForBCIBasedOnSimpleFeatures},
    pdfauthor={Frida Heskebeck},
    }
\usepackage[nameinlink,capitalise]{cleveref}
\crefname{equation}{}{}

\newcommand{\myMethod}[0]{\emph{Transfer Performance Predictor }}
\newcommand{\myMethodShort}[1]{\emph{TPP{#1}}}

\DeclareMathOperator*{\argmin}{arg\,min}

\begin{document}
\title{Source Data Selection for Brain-Computer Interfaces based on Simple Features\footnote{This work was partially supported by the Wallenberg AI, Autonomous Systems and Software Program (WASP) funded by the Knut and Alice Wallenberg Foundation. The authors are members of the ELLIIT Strategic Research Area at Lund University.}}

\author{Frida Heskebeck\thanks{F. Heskebeck is with the Department of Automatic Control, Lund University, Lund, Sweden}, Carolina Bergeling\thanks{C. Bergeling is with the Department of Mathematics and Natural Sciences, Blekinge Tekniska Högskola, Karlskrona, Sweden}, Bo Bernhardsson\thanks{B. Bernhardsson is with the Department of Automatic Control, Lund University, Lund, Sweden}
}
\markboth{IEEE Transactions on Human-Machine Systems}%
{}
\date{2024}

\maketitle
\noindent
\emph{\Large{This work has been submitted to the IEEE for possible publication. Copyright may be transferred without notice, after which this version may no longer be accessible.}}

\begin{abstract}
\noindent
This paper demonstrates that simple features available during the calibration of a brain-computer interface can be utilized for source data selection to improve the performance of the brain-computer interface for a new target user through transfer learning. To support this, a public motor imagery dataset is used for analysis, and a method called the \myMethod method is presented. The simple features are based on the covariance matrices of the data and the Riemannian distance between them. The \myMethod method outperforms other source data selection methods as it selects source data that gives a better transfer learning performance for the target users. 

\noindent
\textbf{Keywords: 
Brain-Computer Interface,
Calibration,
Cross Subject,
Machine Learning, 
Riemannian Geometry,
Source Data Selection,
Transfer Learning,
}
\end{abstract}

\section{Introduction}\label{sec:introduction}
Before a brain-computer interface (BCI) can be used by a new user, referred to as the target user, the system needs to be calibrated. Calibration involves collecting training data from the target user and training the underlying machine learning algorithm, a process that takes a long time and is tiresome for the user. This calibration requirement is one of the main obstacles for everyday ready-to-use BCIs \cite{lotte_signal_2015,heskebeck_calibration_2023}. 

One approach to reducing the calibration time is to utilize data from another user, called source data, with transfer learning to train the underlying machine learning algorithm for the target user. Then, less training data from the target user needs to be collected, significantly reducing the calibration time \cite{nam_transfer_2018, wan_review_2021}. However, it has been shown that transfer learning does not always work for any combination of source and target data in BCI applications \cite{coelho_rodrigues_when_2019}. Thus, it is desirable to identify a combination of source and target data that gives the target user a good transfer learning performance. The success of transfer learning largely depends on the so-called intra-subject performance—i.e., the BCI's performance when only the target user's data is used—information that is not available when a new target user begins using the BCI \cite{coelho_rodrigues_when_2019}. This leaves us with the research question of quickly identifying good source data for a target user, only using features that are available during the calibration of the BCI. 

This paper demonstrates that simple features of the target data, available during the calibration of the BCI system, can be used to select suitable source data for transfer learning. To validate this approach, the \myMethod (\myMethodShort) method is presented for source data selection, which improves the BCI performance compared to several baseline methods for source data selection. As always in processing data for BCIs, details in the processing can be changed to improve the performance for a specific user or scenario of the BCI. To avoid getting lost in these case-to-case details, this paper uses a public dataset with data from many users, a fixed processing pipeline, and one transfer learning method.

\cref{sec:related_work} presents other research on source data identification for BCIs. \cref{sec:background} provides some background to BCI systems and the transfer learning method for readers unfamiliar with BCIs. \cref{sec:material_method} presents the dataset and methods used in the paper. \cref{sec:results} presents and discusses the results. \cref{sec:discussion} puts the results into the bigger picture of source data selection for BCI calibration and outlines future research. Finally, \cref{sec:conclusions} summarizes the paper.

\section{Related work}\label{sec:related_work}
\noindent
Selecting suitable source data for transfer learning is not a novel idea in itself; it has been studied before. This paper’s novelty lies in its approach to source data selection. There are many approaches to source data selection in the literature. One approach is to evaluate the transfer learning performance for each combination of source and target users and use the results to determine if the users are compatible \cite{wei_multi-source_2023}. This corresponds to the baseline source data selection method called \emph{Oracle} in this paper. An alternative approach is to compare the distance between the class means for the target and source data \cite{orihara_active_2023,wu_multi-source_2023}. This approach is similar to the baseline source data selection method called \emph{Distance} in this paper. 
Yet another approach is to study the power spectral densities for resting-state EEG \cite{jeon_domain_2019}. A different approach used in some papers, \cite{xu_selective_2021,li_transfer_2021}, is to use the so-called sequential forward floating search method with different improvement measures for the method. The sequential forward floating search finds suitable source data by iteratively including and excluding source data until a set of source data giving a good BCI performance for the target user has been found. In one paper, \cite{xu_selective_2021}, an improvement measure based on the scatter matrices of the class means is used. In another study, \cite{li_transfer_2021}, an improvement measure based on the transfer learning accuracy is used. All the papers mentioned above use source selection as part of a classification pipeline and aim to improve classification accuracy. However, they are generally not concerned about the calibration time of the BCI, and some of the methods take a long time to identify the source user, making them unusable in an everyday BCI setting.

\section{Background}\label{sec:background}
\noindent
This section provides background to electroencephalography (EEG) based brain-computer interfaces (BCI), how BCI data can be classified, and the Riemannian Procrustes Analysis (RPA) transfer learning method used in this paper. Readers familiar with these topics can skip to \cref{sec:material_method}.

\subsection{EEG-based motor imagery BCIs}
\noindent
BCIs enable the use of thought commands as input to computers instead of other input modalities, such as moving a computer mouse to control the computer. Most BCIs use EEG to measure brain activity from electrodes placed on the scalp of the user. EEG signals are noisy, and extracting the thought commands from EEG signals requires advanced preprocessing of the signals and machine learning algorithms to decipher the meaning. The data in this paper is from the so-called motor imagery (MI) BCI paradigm, which means that the user imagines moving a body part, and the aim is to decode which body part the user imagined moving. In a BCI setting, a predefined set of body movements is used, where each movement corresponds to a specific command, such as ‘right hand $\rightarrow$ move cursor right’ when controlling a screen cursor. A thorough introduction to BCIs is given in \cite{heskebeck_calibration_2023,nam_braincomputer_2018-ch1}.

\subsection{Classification of BCI data using Riemannian Geometry}
\noindent
There are many approaches to decoding and classifying BCI data. In recent years, a common approach for MI data has been to use the covariance matrix for a trial's EEG data as a feature representation and then use Riemannian geometry when comparing covariance matrices from different trials. The term trial refers to the experimental session where the user imagines the movement of one body part. 
A thorough explanation of Riemannian approaches in BCIs is given in \cite{yger_riemannian_2017}. Here, a summary is presented to provide context for this paper. 

The covariance matrix, $C_i$, is found from the EEG data during an MI trial and is a feature representing the imagined movement of one body part during trial $i$. A dataset consists of labeled covariance matrices $(C_i,y_i)$ where $y_i \in \left\{ 1,..., K \right\}$ is the label indicating which of the $K$ predefined body parts was imagined moved during trial $i$. Let $X_i\in \mathbb{R}^{ch\times s}$ be the EEG data for trial $i$ recorded with $ch$ electrodes for $s$ time samples. The covariance matrix for that trial is calculated as 
\begin{equation*}
    C_i=\frac{1}{s}X_iX_i^\top.
\end{equation*}

Covariance matrices are symmetric positive definite (SPD) and live on the Riemannian space of SPD matrices $\mathcal{P}_n$. Using the Affine Invariant Riemannian Metric (AIRM), the distance between two covariance matrices on the manifold $\mathcal{P}_n$ is 
\begin{equation*}
    \delta_R(C_i,C_j) = || \log(C_i^{-1/2}C_jC_i^{-1/2})||_F,
\end{equation*}
which hereafter is called the Riemannian distance. Using the Riemannian distance, the geometric mean $M$ (a covariance matrix) of the set $\mathcal{C}$ of $L$ covariance matrices is 
\begin{equation}\label{eq:riemann_mean}
    M =\argmin_{X\in \mathcal{P}_n} \sum_{C_i\in \mathcal{C}} \delta_R^2(X,C_i).
\end{equation}
Assuming that the data belongs to $K$ different classes, we can calculate the mean for each class, $M_k$, using \cref{eq:riemann_mean} and the covariance matrices with label $k$, $\mathcal{C}_k =\left\{ C_i\, |\, y_i = k\right\}$, as 
\begin{equation}\label{eq:riemann_mean_class_k}
    M_k =\argmin_{X\in \mathcal{P}_n} \sum_{C_i\in \mathcal{C}_k} \delta_R^2(X,C_i).
\end{equation}
The dispersion, $d$, of the data belonging to class $k$, around its geometric class mean $M_k$ will be defined as 
\begin{equation}\label{eq:rieman_dispersion}
    d_k = \frac{1}{|\mathcal{C}_k|} \sum_{C_i\in \mathcal{C}_k} \delta_R^2\left(M_k,C_i \right),
\end{equation}
where $|\mathcal{C}_k|$ is the number of covariance matrices in the set $\mathcal{C}_k$.

Using this Riemannian framework, a common classification algorithm is the minimum distance to mean (MDM) classification where a new sample $C_{\text{new}}$ is classified as the class for the mean $M_k$ it is closest to,
\begin{equation*}
    y_\text{new} = \argmin_{k\in K} \, \delta_R^2\left(M_k,C_\text{new}\right).
\end{equation*}

\subsection{Transfer learning with the RPA method}
\noindent
The Riemannian Procrustes Analysis (RPA) is introduced and explained in detail in \cite{yger_riemannian_2017}. The code is implemented in the pyRiemann Python package \cite{pyriemann}. Here, a summary is presented to provide context for this paper. 

Assuming that the data for a trial $i$ is represented by the covariance matrix $C_i$ with label $y_i$, as described above. The set of \emph{source} data $\mathcal{S}$ with $L^\mathcal{S}$ covariance matrices and \emph{target} data $\mathcal{T}$ with $L^\mathcal{T}$ covariance matrices are
\begin{align*}
\begin{split}    
    \mathcal{S} &= \left\{(C_i^\mathcal{S}, y_i^\mathcal{S}) \text{ for } i = 1,...,L^\mathcal{S} \right\}, \\
    \mathcal{T} &= \left\{(C_i^\mathcal{T}, y_i^\mathcal{T}) \text{ for } i = 1,...,L^\mathcal{T} \right\}.
\end{split} 
\end{align*}
For the source and target domain data separately, the mean $M$ for all the data and the mean for each class $M_k$ is found with \cref{eq:riemann_mean} respectively \cref{eq:riemann_mean_class_k} as $M^{\mathcal{S}}$, $M^{\mathcal{S}}_k$, $M^{\mathcal{T}}$, and $M^{\mathcal{T}}_k$.
Similarly, the dispersion $d$ for all data and the dispersion $d_k$ for each class $k$ is found with \cref{eq:rieman_dispersion} as $d^{\mathcal{S}}$, $d^{\mathcal{S}}_k$, $d^{\mathcal{T}}$, and $d^{\mathcal{T}}_k$.
The means and dispersions of the data are used in the RPA method for transfer learning and in this paper as features for the \myMethod method.

Transfer learning with the RPA method is done in three steps \cite{yger_riemannian_2017}: 
\begin{enumerate}
    \item Re-center the data so that $M^\mathcal{S}$ and $M^\mathcal{T}$ become identity.
    \item Equalize the dispersions $d^\mathcal{S}$ and $d^\mathcal{T}$ by moving the matrices along their geodesic to the identity matrix. 
    \item Rotate the data around the geometric mean (identity) to align the class means $M_k^\mathcal{S}$ and $M_k^\mathcal{T}$ for each class $k$.
\end{enumerate}
The first two steps can be done on unlabeled data, the unsupervised part of the RPA method, but the last step requires labeled data, the supervised part of the RPA method. After transfer learning, the source data and target data are aligned in the sense that the data from each class of the two datasets are matched.

\section{Material and Method} \label{sec:material_method}
\noindent
This paper uses a motor imagery (MI) dataset where the users in each trial imagine moving either their hands or feet. The task is to determine which body part the user imagined moving \cite{schalk_bci2000_2004}. Each trial's covariance matrix is used as a feature, and minimum distance to mean (MDM) classification is used for the MI classification. The transfer learning of two users' data, the target and source user, is done with the Riemannian Procrustes Analysis (RPA) transfer learning method \cite{rodrigues_riemannian_2019}. 
In a follow-up paper \cite{coelho_rodrigues_when_2019}, the authors conclude that the most predictive feature for determining whether transfer learning with the RPA method will be successful is the target user's intra-subject MI classification accuracy. 
However, as they note, this feature is not available in real calibration settings where no prior information about the new target user is known, raising the research question of how to identify suitable source data for a target user \cite{coelho_rodrigues_when_2019}. 

This paper presents the \myMethod (\myMethodShort) method for source data selection in a BCI transfer learning setting. The method predicts the transfer learning performance for a target and source user based on simple data features that are available during the calibration of a BCI. 
The method outperforms other source data selection methods and uses features available during a BCI's calibration, thus taking a step toward improving BCI calibration.

The following subsections present details about the dataset and the data processing, the MI classification and the transfer learning, the \myMethodShort{} method, and how the source data selection methods are compared. The source code for the project is available online\footnote{\url{https://gitlab.control.lth.se/FridaH/source-data-selection-public}}.

\subsection{Dataset and Preprocessing}
\noindent
The public Physionet dataset, \cite{schalk_bci2000_2004}, is used for the analysis in this paper. The dataset includes labeled EEG data for experiments where the users conducted a hands vs. feet motor imaginary task, a right-hand vs. left-hand motor imaginary task, a hands vs. feet motor movement task, and a right-hand vs. left-hand motor movement task. Both motor imaginary tasks were analyzed, but only the results for the hands vs. feet motor imaginary task is presented in this paper as the result is similar for both tasks. The dataset includes data from 109 users, and each user has data from about 45 trials of the hands vs. feet motor imaginary task.

The MNE Python toolbox is used for the preprocessing of the data \cite{MNE_paper,MNE_toolbox}. The nine EEG channels F3, Fz, F4, C3, Cz, C4, P3, Pz, and P4 are used as they cover the motor cortex of the brain \cite{heskebeck_calibration_2023}. The EEG data is bandpass filtered in the range 7-35 Hz as it covers the frequency bands where interesting EEG features are expected to appear \cite{heskebeck_calibration_2023}. Each trial is from 1 sec after cue onset until 2 sec after cue onset to avoid any initial transients from the stimuli onset. The $9\times9$ covariance matrix for each trial is used to represent the trial's data. 
See the provided code\footnotemark[1] for implementation details. Only users who had an intra-subject MI classification accuracy above 65\% (98 users) were used in the paper, as it is known from previous research that the RPA algorithm only works for users with sufficient intra-subject performance \cite{coelho_rodrigues_when_2019}.

\subsection{MI classification with transfer learning}
\noindent
The MI task is classified with covariance matrices as features using the MDM classification based on the Riemannian geometry for SPD matrices with the Affine Invariant Riemannian Metric (AIRM) \cite{yger_riemannian_2017}. The transfer learning is done with the RPA method \cite{rodrigues_riemannian_2019}. Target domain data refers to data from the new user where a limited amount of data and knowledge about the data exist. Source domain data refers to existing data from other users that can be used for transfer learning. 
The analysis in this paper assumes that the mean covariance matrix for each MI class and the dispersion around these class means are known for the target data. How much training data is needed to estimate these features is a topic for future research. 

For each combination of source--target users, a subset of the data from the target user, called target training data, and all data from the source user are used for the RPA transfer learning method. The target training data is the data that needs to be collected during the calibration phase of the BCI. The remaining data from the target user is used as target test data. We omit the second step of the RPA method, as it was found to distort the data more than improve it. This distortion is likely due to poor estimation of the dispersion in the target data, given the limited amount of available data. With more target data and, consequently, a more accurate dispersion estimate, this step should be included in the RPA method.
After transfer learning, the MDM classifier is trained with the source training data, not the target training data, and tested on the target test data in order to ensure that it is the cross-subject transfer learning performance that is evaluated and not the intra-subject performance. The resulting cross-subject MI classification accuracy, also called transfer learning accuracy or cross-subject performance, is stored. 

The above-described procedure for splitting the target data into training and test data, doing RPA transfer learning, and then MDM classification is hereafter called MI classification.
The MI classification is done with 5-fold cross-validation, meaning that the above procedure is repeated five times for each source--target combination, with different splits of target training/test data in each fold. The average cross-subject MI classification accuracy over the 5-fold cross-validation is shown in \cref{fig:accuracy}.
For the case where the source and target user are the same (the highlighted cells in \cref{fig:accuracy}), the intra-subject accuracy is shown. In this case, no transfer learning is done, and the target training data is used for the MDM classification. The target test data is still used as test data. 

\subsection{The \emph{\myMethod} \emph{(\myMethodShort)} method}\label{subsec:method:myMethod}
\noindent
This paper's main point is to show that simple features describing the source and target data can be used to select source data for BCI transfer learning. This is concretely shown with the \myMethod (\myMethodShort) method, which outperforms other source data selection methods described in the following subsection. 
The \myMethodShort{} method predicts the cross-subject MI classification accuracy, also called transfer learning accuracy. In other words, the \myMethodShort{} method predicts the color for the source--target combinations in \cref{fig:accuracy}. For source data selection, the source data with the highest predicted cross-subject MI classification accuracy for the target user is selected. 

The \myMethodShort{} method uses 18 features for the prediction of cross-subject performance. The features are listed in \cref{tab:features} for target data $\mathcal{T}$, source data $\mathcal{S}$, MI class~1 and MI class~2. The features are categorized into distances, dispersions, accuracies, and differences and are mainly based on the distances between the class means for the source and target data. The observant reader will notice that most of these features are the same as those used in the RPA transfer learning algorithm and those used in the MDM classifier; this is not a coincidence. These features are selected since they are easy to find with limited training data and are important since both the RPA and MDM classification algorithms are based on versions of these features. The features are calculated after the data has been recentered but before the data has been rotated in the RPA transfer learning algorithm.

\begin{table}
    \caption{The features for the \myMethod method are categorized into distances, dispersions, accuracies, and differences.}
    \label{tab:features}
    \renewcommand{\arraystretch}{1.3}
    \begin{tabular}{llll}
    \hline
         \textbf{Distances} & \textbf{Dispersions} \\
         \hline
         $\delta_R^2(M_1^\mathcal{T},M_2^\mathcal{T})$ & $d^{\mathcal{T}}$ \\
         $\delta_R^2(M_1^\mathcal{S},M_2^\mathcal{S})$ & $d^{\mathcal{T}}_1$ \\
         $\delta_R^2(M_1^\mathcal{T},M_1^\mathcal{S})$ & $d^{\mathcal{T}}_2$ \\
         $\delta_R^2(M_1^\mathcal{T},M_2^\mathcal{S})$ & $d^{\mathcal{S}}$  \\
         $\delta_R^2(M_2^\mathcal{T},M_1^\mathcal{S})$ & $d^{\mathcal{S}}_1$ \\
         $\delta_R^2(M_2^\mathcal{T},M_2^\mathcal{S})$& $d^{\mathcal{S}}_2$ \\
         & \\
         \hline
         \textbf{Accuracies} & \textbf{Differences} \\
         \hline
         Source data intra-subject & $\delta_R^2(M_1^\mathcal{T},M_2^\mathcal{T}) - \delta_R^2(M_1^\mathcal{S},M_2^\mathcal{S})$ \\
         & $\delta_R^2(M_1^\mathcal{T},M_2^\mathcal{S}) - \delta_R^2(M_1^\mathcal{T},M_1^\mathcal{S})$ \\
         & $\delta_R^2(M_2^\mathcal{T},M_1^\mathcal{S}) - \delta_R^2(M_2^\mathcal{T},M_2^\mathcal{S})$ \\
         & $d^{\mathcal{S}}_1 -d^{\mathcal{T}}_1$ \\
         & $d^{\mathcal{S}}_2 - d^{\mathcal{T}}_2$ \\
    \end{tabular}
\end{table}

Scikit-learn, \cite{scikit-learn}, is used for the machine learning implementation. The pipeline for the \myMethodShort{} method is Scikit-learn's robust scaler followed by a neural network with two hidden layers (50, 50) with the relu activation function; see the provided code\footnotemark[1] for implementation details. Other pipelines for the \myMethodShort{} method were also tested and performed well. This one was selected for its consistent performance across all analyzed datasets (only the results from the feet vs. hands dataset are presented in the paper, as previously explained).

The result is verified by leave-k-groups-out cross-validation. In each fold, some users are used as test users, and their data is excluded from the training set. 
The test data is used to evaluate the performance of the prediction and, in particular, the performance of the BCI if the \myMethodShort{} method is used for source data selection compared to other source data selection methods (described in the next section). Each user is used as a test target in one of the ten folds.

\subsection{Source data selection}\label{subsec:method:source_selection}
\noindent
As explained before, the overall point of this paper is to show that simple features available during the calibration of a BCI system can be used to select source data for transfer learning, which can streamline the calibration of BCIs. The source data selection task is to identify the source user that gives a target user the highest cross-subject performance. Looking at \cref{fig:accuracy}, which shows the transfer learning accuracy for all combinations of source and target users, and assuming that the target user is at row $i$, the source data selection task is to predict which source user gives the darkest blue value for row~$i$. 

The compared methods for source data selection in this paper are: 
\begin{itemize}
    \item \textbf{\emph{Intra-subject}} -- No source data is used, but the intra-subject performance is. The intra-subject performance is the accuracy of MI classification if only the target training data is used. This corresponds to the BCI's performance if no transfer learning is done and is a baseline for what a source selection method should achieve in performance.
    \item \textbf{\emph{Random source}} -- The source data is selected at random. A useful source data selection method should outperform the Random source method.
    \item \textbf{\emph{Distance}} -- The source data is selected as the source data with class means closest to the target data. In other words, the source $\mathcal{S}$ is selected as $$\mathcal{S} = \argmin_{\mathcal{S}} \frac{\delta_R^2(M_1^\mathcal{T},M_1^\mathcal{S}) + \delta_R^2(M_2^\mathcal{T},M_2^\mathcal{S})}{2}$$ This method is similar to what was done in \cite{orihara_active_2023}. The distances is divided by two in order to be the mean distance.
    \item \textbf{\emph{Best source}} -- The source data is selected as the source user with the highest intra-subject accuracy. This method is based on the assumption that the user who, in some sense, is the ``best'' should be the best to use as source data.
    \item \textbf{\emph{Best teacher}} -- The source data is selected as the source user that is the best for most other target users. This corresponds to selecting the source user to the left in \cref{fig:accuracy} as the source user.
    \item \textbf{\emph{Max of methods}} -- The combined version of the Best source, Best teacher, and \myMethodShort{} source data selection methods. Each included method suggests a source user and the one with best transfer learning performance is selected as the source user.
    \item \textbf{\myMethod (\myMethodShort)} -- The presented method in this paper. It uses simple features available during the calibration to predict the cross-subject performance and selects the source user with the highest predicted performance; see details in \cref{subsec:method:myMethod}. The method is abbreviated TransPerfPred in \cref{fig:baseline_comparision}.
    \item \textbf{\emph{Oracle}} -- The source data is selected as the best source data for each target user. It represents the best possible performance for the BCI. This method is similar to what they did in \cite{wei_multi-source_2023} but is not a viable option during the calibration of a BCI as it takes too long time to evaluate all source users to find the best one. 
\end{itemize}

As mentioned above, evaluating all available source data with the target (the \emph{Oracle} method) is impractical during BCI calibration, as it would be too time-consuming with a large amount of source data. This is primarily because the rotation step in the RPA transfer learning algorithm is time-intensive, requiring solving an optimization problem in the Riemannian space of SPD matrices to find the rotation matrix for each source-target combination \cite{rodrigues_riemannian_2019}. However, it is possible to evaluate a few source data candidates and then select the best among them.
Assuming that six source candidates can be evaluated, it means that the source selection methods \emph{Distance}, \emph{Random source}, \emph{Best source}, \emph{Best teacher}, and \myMethodShort{} select six source candidates. For the \emph{Max of methods} method, which is the combination of three source selection methods, it means that each of the included methods selects two source candidates to get six candidates in total. For the \emph{Oracle} and \emph{Intra-subject} methods, there is always only one candidate. The result from this analysis is shown in \cref{fig:baseline_comparision}. In \cref{fig:baseline_comparision_nbr_candidates}, the methods' performance is compared to the \emph{Oracle} method for different numbers of source data candidates.

\section{Results and Discussion}\label{sec:results}
\noindent
This section presents the results and discusses the MI classification using transfer learning and the source data selection methods. The following section, \cref{sec:discussion}, contextualizes the findings within the broader research question of source data selection and outlines potential future research directions.

\subsection{MI classification with transfer learning}
\begin{figure}[t!]
    \centering
    \includegraphics[width=1\linewidth]{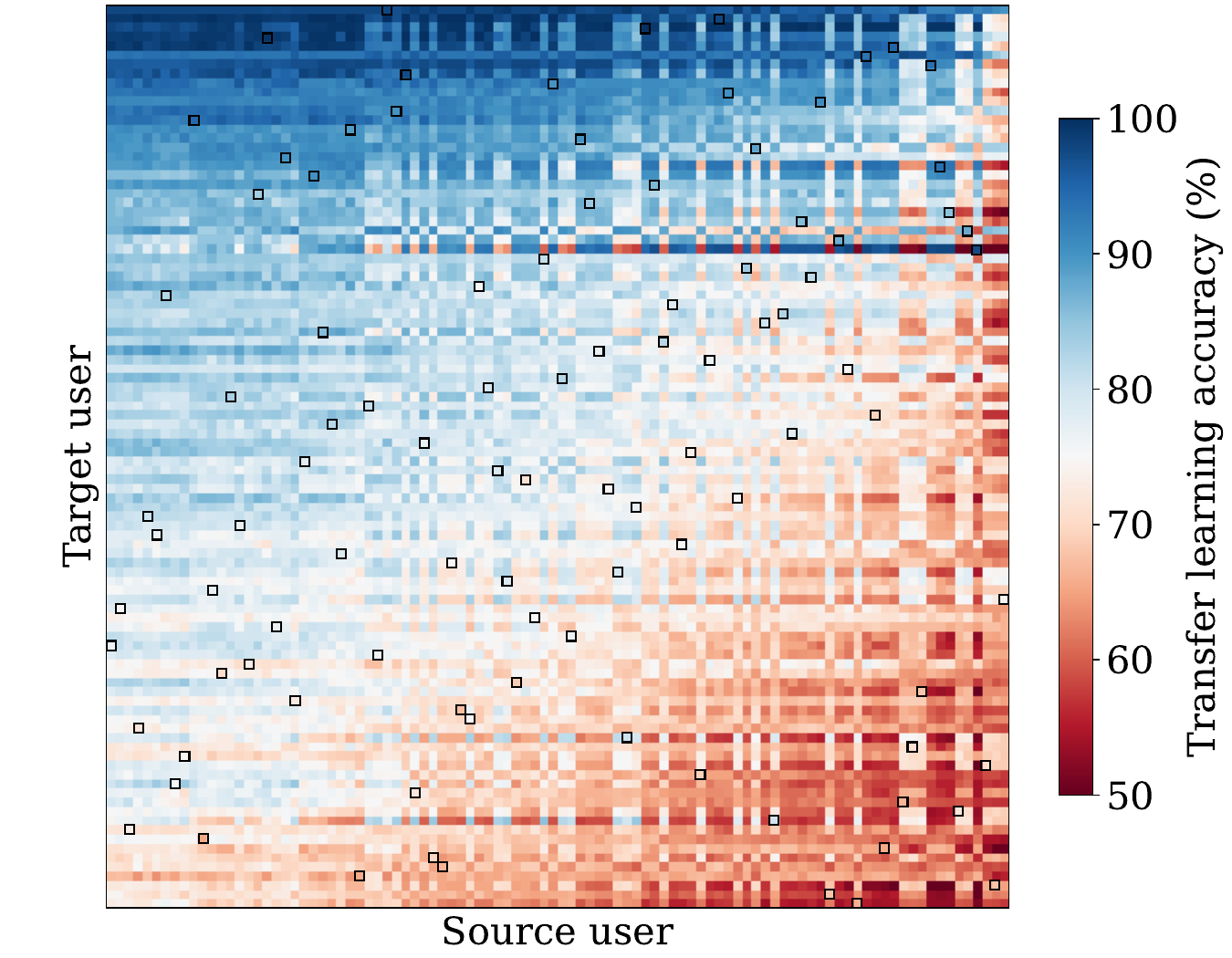}
    \caption{Accuracy of MI classification after using the RPA transfer learning method, average over 5-folds. The rows are target users, and the columns are source users. The colors correspond to the cross-subject MI classification accuracy for each combination of target and source users. The highlighted cells are the intra-subject accuracies. For an arbitrary row, we can see that the transfer learning accuracy differs a lot depending on which source data is used, which is the reason why source data selection for BCIs is important. The rows are sorted in decreasing order of the row sum, and columns are sorted in decreasing order of the column sum. An interpretation of this sorting is that target users that benefit from many source users are found at the top (good students), and source users that are beneficial for many target users are found to the left (good teachers). 
    }
    \label{fig:accuracy}
\end{figure}

\noindent
The matrix in \cref{fig:accuracy} shows the average 5-fold cross-validation results for transfer learning MI classification for all combinations of source--target data. Each row in the matrix corresponds to a target user, and each column corresponds to a source user. The matrix color at index $i,j$ reflects the cross-subject MI classification accuracy for combining the target user at row $i$ and the source user at column $j$. Cross-subject MI classification accuracy, or transfer learning accuracy, means the classification accuracy of target test data when the MI classifier is trained using the target training data and source data after transfer learning. White color corresponds to a 75\% accuracy, blue color to higher than 75\% accuracy, and red color to lower than 75\% accuracy as shown by the colorbar to the right of the matrix. An accuracy of 100\% (dark blue) is the maximum, and an accuracy of 50\% (dark red) corresponds to the chance level for the MI classification task. 

The highlighted cells correspond to indices where the target and source users are the same. The color in these cells is the intra-subject MI classification for that user. Intra-subject MI classification means the classification accuracy of target test data when the MI classifier is trained on only target training data without transfer learning. Note that the highlighted cells are not on the main diagonal since the columns and rows are sorted separately, as explained in the next paragraph.

The matrix's rows are sorted in descending order of row summation. This means a target user with high transfer learning accuracy for many source users is found at the top, and a user with low transfer learning accuracy for most source users is found at the bottom. The matrix's columns are sorted similarly, in descending order of column summations. This means that a user whose data is beneficial for many target users is found to the left of the matrix, and a user whose data is disadvantageous for many target users is found to the right of the matrix. Thus, the order of users is different in the rows and columns, which is why the highlighted cells, corresponding to indices where the source and target user are the same, are not found on the main diagonal. 

There are a few observations to highlight from \cref{fig:accuracy}. The first is if we interpret the target users as students and the source users as teachers, an assumption would be that a good teacher, i.e., someone who can teach any student, would also be a good student, i.e., someone who can learn from any teacher. However, this is not the case since the best teachers, placed to the left in the matrix, are found to be ``middle-performing'' students. We can conclude this from the highlighted cells for these users, which are the furthest to the left but in the middle of the rows. Another observation from \cref{fig:accuracy} is the cluster of highlighted cells in the bottom right corner. From these, we can generally say that a bad student is often a bad teacher. Besides this, there is no clear clustering of the source and target users for this matrix sorting. 
A final observation is that the good students at the top of the matrix generally have high intra-subject accuracies, as seen from the highlighted cells, and the worst students at the bottom of the matrix have low intra-subject accuracies. This is expected from the results of \cite{coelho_rodrigues_when_2019}, which said that the intra-subject performance is the most telling feature for the transfer learning performance.

Looking at a single row, it is clear that a target user benefits from source data selection for transfer learning since some source data have low transfer learning accuracy and some have high transfer learning accuracy. For the very best students, at the top of the matrix, it doesn't matter as much which source data is selected; any source data performs well. But for a target user in the middle of the matrix, we see that finding a good source user could give a transfer learning accuracy up to 80\% while a bad source user could give a transfer learning accuracy down to 60\%, which makes a difference for the final performance of the BCI. Even the worst performing student, at the lowest row, benefits a lot from source data selection, and it may be for these middle- and less-performing users that source selection is the most important.

\subsection{Source data selection}
\begin{figure}[t!]
    \centering
    \includegraphics[width=1\linewidth]{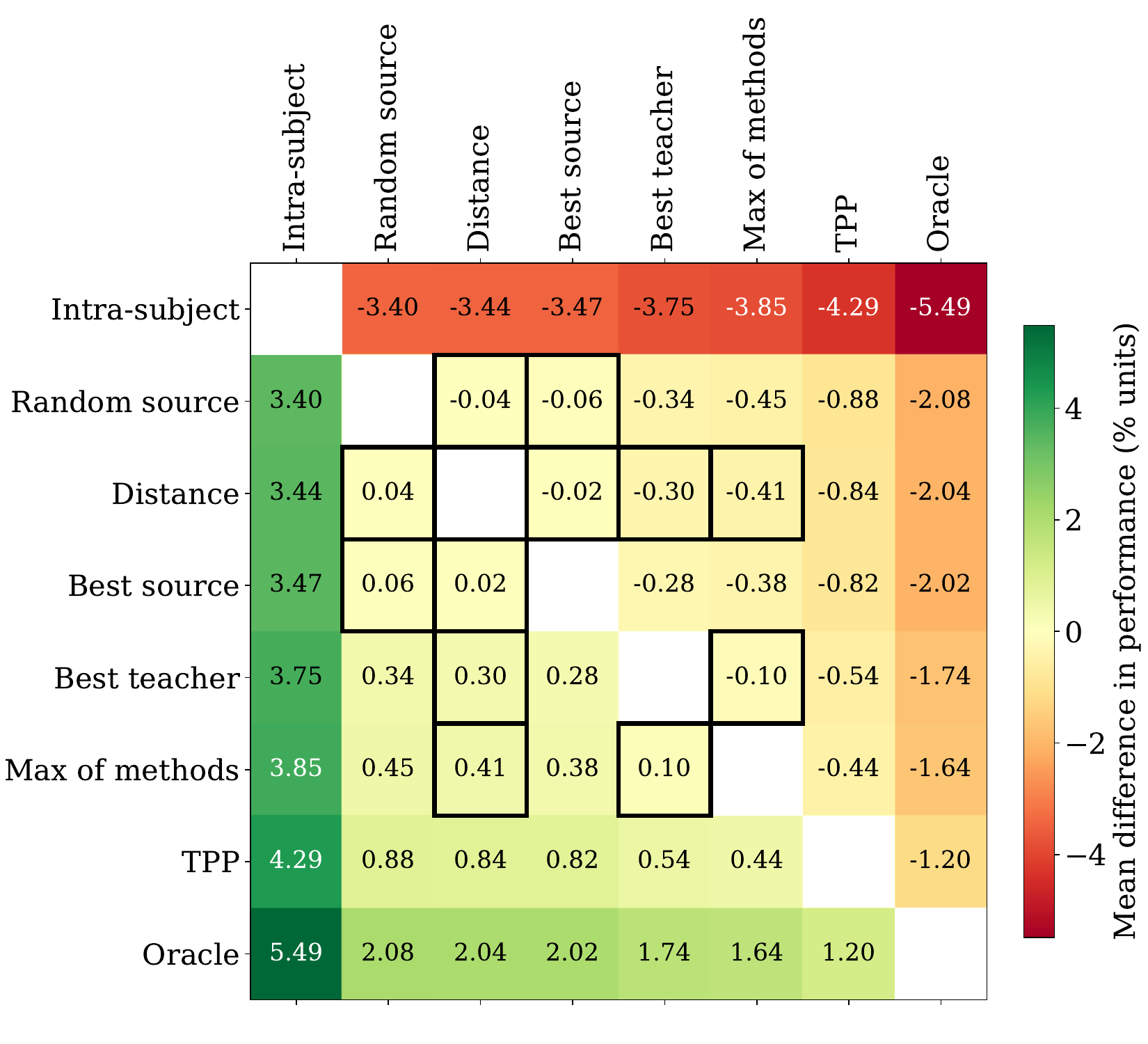}
    \caption{Comparision of different source data selection methods. The colors and numbers represent the mean difference in transfer learning accuracy between the method in the row and the method in the column. The matrix is thus skew-symmetric. A higher number means that the method in the row, on average, selects source data that gives each target user a higher transfer learning accuracy than the method in the column. The highlighted cells indicate methods where the difference between the transfer learning accuracies for each target user is not statistically different from 0, meaning that one cannot say that one method is better. As expected, the \emph{Oracle} method is the best.
    }
    \label{fig:baseline_comparision}
\end{figure}

\noindent
The matrix in \cref{fig:baseline_comparision} shows a comparison of the different source data selection methods. The numbers and colors represent the mean difference in transfer learning accuracy when the source data is selected with the method in the row compared to the method in the column. The difference between the transfer learning accuracies for each target user is calculated before the mean is calculated. 
For example, the mean difference between the \emph{Oracle} and \emph{Random source} methods, located in the lower-left corner of the matrix, is 2.08. This indicates that the \emph{Oracle} method identifies a source user that yields, on average, 2.08 percentage points higher transfer learning accuracy compared to the source user selected by the \emph{Random source} method. The matrix is skew-symmetric since the value in the matrix shows the difference between the method in the row and the method in the column, which has the opposite sign when the row and column change. 
As described in \cref{subsec:method:source_selection}, the result in \cref{fig:baseline_comparision} is from when the methods select six candidates for source data. All six candidates are evaluated, and the best of these is selected as the source data.

The highlighted cells indicate comparisons between methods where the difference in transfer learning accuracy for the target users is not significantly different from zero using the Wilcoxon signed-rank test with a p-value of 0.05. This means that for the highlighted cells, one cannot say that one method is better than the other. Looking at the comparison of the \emph{Distance} method to the \emph{Max of methods} method, the cell is highlighted but has a value of -0.41. This means that, on average, the difference is -0.41 when comparing these methods. However, when looking at the difference for each target user, most differences are around 0 with a median value of 0. The -0.41 value comes from a few target users where the \emph{Max of methods} method has found a significantly better source user than the \emph{Distance} method, which bumps up the mean value. Thus, based on the statistical test, we cannot say that either method found source data that performed better than the other, even though the value was -0.41. The median value is not shown here but is zero or close to zero for the highlighted cells.

One observation from these results is that the \emph{Intra-subject method} for source selection is the worst compared to all methods. This is due to the minimal training data for the MI classifier. The low performance of the method implies that in a calibration setting where there is only a limited amount of target training data available, using transfer learning is generally better than using only the target training data. The difference between the \emph{Intra-subject} method and the \emph{Oracle} method tells us how much there is to gain from doing transfer learning.

A second observation is when comparing the \emph{Oracle} method with the other methods, a value closer to zero means that the compared method is closer to finding a source user that performs as well as the \emph{Oracle} source data. The numbers show that the \myMethod (\myMethodShort) method performs the closest to the \emph{Oracle} method.

A third observation is that the \emph{Random source} method puts a baseline for when a method is better than chance. The highlighted cells show that the \emph{Distance} and \emph{Best source} methods are not significantly better than the \emph{Random source} method. The \emph{Best source} method selects source users with the best intra-subject MI classification accuracies. The fact that this method is not better than chance indicates that it is more important that the source and target users are similar to each other than the source user being the ``best'' BCI user, i.e., having good intra-subject MI classification. Measuring the similarity of users is a fascinating topic and will be explored in future research.

A fourth observation is that the \emph{Best teacher} method is better than chance, which is reasonable since it selects source users that benefit most target users. However, it is not significantly better than the \emph{Distance} method, as indicated by the highlighted cell, even though the mean difference of 0.36 is relatively high between these methods. The reason for this is similar to the example previously; there are some target users for whom the \emph{Best teacher} method finds a better source user than the \emph{Distance} method, bumping up the mean value, but for most users, the methods are equally good.

A final observation is that the \myMethodShort{} method is significantly better than all other methods except for the \emph{Oracle} method, which is, as expected, better. Since the \myMethodShort{} method outperforms the other methods, we can conclude that using simple features to select source data is a working and promising approach. The \myMethodShort{} method does not find the optimal source data but is a step towards improving the source selection for transfer learning during BCI calibration.

\begin{figure}[t!]
    \centering
    \includegraphics[width=1\linewidth]{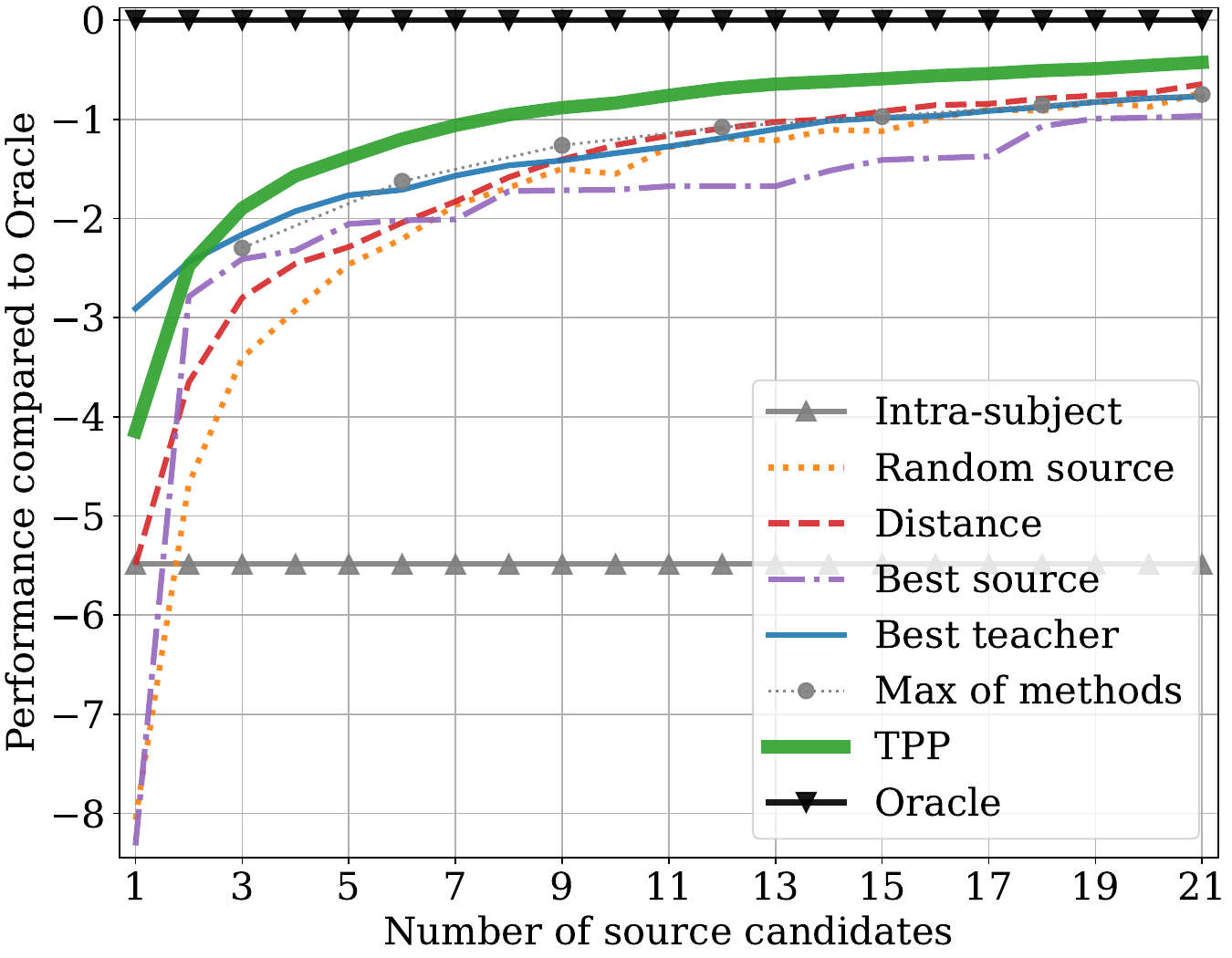}
    \caption{Comparision of the source data selection methods for different numbers of source data candidates. The y-axis shows the method's performance compared to the \emph{Oracle} performance. The x-axis shows the number of candidates suggested by each method for source selection. The figure shows the same numbers for six source candidates as the rightmost column in \cref{fig:baseline_comparision}. The more candidates each method can suggest, the closer to \emph{Oracle} performance. The \myMethodShort{} method outperforms the other for three or more suggested candidates. 
    }
    \label{fig:baseline_comparision_nbr_candidates}
\end{figure}
\cref{fig:baseline_comparision_nbr_candidates} shows the methods' performance compared to the \emph{Oracle} method for different numbers of source candidates. The rightmost column in \cref{fig:baseline_comparision} corresponds to \cref{fig:baseline_comparision_nbr_candidates} at six source candidates. For the \emph{Max of methods} method, only results for multiples of three are shown since the \emph{Max of methods} is a collection of three methods and thus will have source candidates in multiples of three. In other words, when the methods included in the \emph{Max of methods} method suggest three source candidates each, the \emph{Max of methods} method will have nine source candidates to compare and should thus be compared with the other methods when they suggest nine candidates. When looking at \cref{fig:baseline_comparision_nbr_candidates}, one should remember that some methods are not statistically different, as we saw in \cref{fig:baseline_comparision}. However, general conclusions can still be drawn.

From \cref{fig:baseline_comparision_nbr_candidates}, we see that the \myMethodShort{} method outperforms the other methods, except for the \emph{Oracle} method, for all numbers of source candidates higher than three. For most BCI settings, it is reasonable to have time to evaluate three or more source candidates, which means that the \myMethodShort{} method is profitable for most BCI applications. Another observation is that the more source candidates, the better all the methods perform. This is reasonable since selecting more source candidates increases the probability of picking a good one. What is a bit unexpected is that only the \myMethodShort{} method outperforms the \emph{Random source} method even when the methods select many source candidates.

\section{Conclusions and Future Research}\label{sec:discussion}
\noindent
This section discusses how the results relate to the bigger picture of source data selection for the calibration of BCI and outlines future research directions. 

\subsection{MI classification with transfer learning}
\noindent
The most important conclusion and takeaway from the results of the MI classification with transfer learning is that source data selection for BCI is needed since the transfer learning performance differs quite a lot depending on which source data is used for a target user. 
As noted in the Results and Discussion section, the difference between a good and a bad source user is higher for low-performing target users than high-performing target users. Thus, source selection is essential for low-performing target users.

Using the interpretation from the Results and Discussion section that a target user is a student and a source user is a teacher, we can first conclude that if it were the case that a good student is also a good teacher, all highlighted cells in the matrix in Figure~\ref{fig:accuracy} would be placed on the main diagonal of the matrix. As stated above in the Results and Discussion section, the best student (the top row) is a relatively bad teacher, and the best teacher (the left column) is a relatively bad student. This means that the data from a user that benefits from transfer learning from many other users is not necessarily good data to use for transfer learning for other users. This contradicts the intuition of transfer learning for BCIs, as one would expect that if source user A is good for target user B, source user B would be good for target user A. The reason why there is an implication and not an equivalence in the transferability of source and target data is something to study in future research. Related to this is that one would assume that there are clusters of users who benefit from each other in terms of transfer learning. The results of this work can neither confirm nor contradict the existence of clusters and cannot say anything about how to find such clusters. That is also a topic for future research.

\subsection{Features for the \emph{\myMethod} method}
\noindent
As motivated in \cref{subsec:method:myMethod}, the features for the \myMethod (\myMethodShort) method are chosen because they are accessible during calibration of the BCI with a limited amount of target training data and are used by both the RPA transfer learning method and the MDM classification method. 
Since the \myMethodShort{} method outperforms other source data selection methods, we conclude that these simple features can effectively guide source data selection, which is the central aim of this paper.

Looking at the features used, see \cref{tab:features}, one could find more variants of the distance measure than those in the table. However, the distances in \cref{tab:features} represent all possible distances between the source and target datasets since many distances are the same. To be precise, the included distances represent all possible distances between the class means $M_1^\mathcal{T}$,$M_2^\mathcal{T}$,$M_1^\mathcal{S}$, and $M_2^\mathcal{S}$ since for example, $\delta_R^2(M_1^\mathcal{T},M_2^\mathcal{S})$ is the same as $\delta_R^2(M_2^\mathcal{S},M_1^\mathcal{T})$.

At this point, a relevant question is how much target training data is needed to estimate these features. In this paper, these features are assumed to be known, and we do not study how much data is needed to estimate them accurately; it is a topic for future research.

A final note on the feature discussion is that the used dataset has two classes, meaning the target and source data have two mean covariance matrices, one for each class. If the dataset had more than two classes, it would mean more mean covariance matrices, which in turn would mean more features from distance, dispersion, and difference measures. The principle for the \myMethodShort{} method would remain the same. 

\subsection{Source data selection methods}
\noindent
There are a few interesting observations from the source data selection comparison to discuss.  
Firstly, as expected, the \emph{Oracle} method is the best source data selection method. However, in a BCI setting where we aim to reduce the calibration time for a new target user, doing transfer learning for all available source data to identify the one with the best performance takes too long to be a viable approach. This is why source data selection is needed. Apart from the \myMethodShort{} method outperforming the other source data selection methods, the strength of the \myMethodShort{} method is that it uses simple features of the target data that are available during the calibration of the BCI. This means that the \myMethodShort{} method can select a source user without the heavy computations required for transfer learning as in the \emph{Oracle} method.
Secondly, picking a random source for transfer learning is better than doing no transfer learning at all, assuming that no more target training data can be collected. Since we also know that it makes a significant difference which source user is used for transfer learning, this only strengthens the argument that source selection for BCI calibration is crucial.

As mentioned in the results section, all methods perform closer to the \emph{Oracle} method when more candidates for source data selection are selected. This is not too strange if we consider the \emph{Random source} selection method. If the \emph{Random source} method can select nine candidates instead of three, it has a higher probability of selecting a good one. A similar reasoning applies to all methods. When fewer candidates can be selected, more requirements are placed on the source selection method to make a good selection.

An alternative to the \myMethodShort{} method is to train a large neural network capable of handling everything from preprocessing to MI classification, similar to the approaches used in image processing. While such networks are beginning to emerge for BCI data, the current limitation is the lack of sufficient BCI data to train them to the same level of reliability as image processing networks. However, as more BCI data becomes available, these networks will likely become more prevalent, and the need for source selection may diminish in the future.

As noted in the introduction, BCI research involves tuning hundreds of parameters in EEG preprocessing, transfer learning, and MI classification. This paper demonstrates that the \myMethodShort{} method is effective for the studied case, suggesting it could be applicable to other scenarios. However, since all BCI data is unique, case-specific considerations are necessary. Nonetheless, the core principle of the \myMethodShort{} method—predicting transfer learning accuracies from simple features for source data selection—remains relevant across all BCI cases.

\subsection{Future work}
\noindent
In the paper, two main directions for further research have been mentioned. This section summarizes these to facilitate further research for researchers who want to continue working on this. 

The first research direction concerns how much training data is needed to accurately estimate the mean covariance matrices. In this paper, it is assumed that the mean covariance matrix for each class in the target data, $M_1^\mathcal{T}$ and $M_2^\mathcal{T}$, is known. The mean covariance matrices are estimated from the target data, and it is important to know how much data is needed to have a good estimate as it determines how much target training data needs to be collected from the target user during the calibration of the BCI. It is a tradeoff between reducing the calibration time by collecting less data and having a sufficiently good estimate of the covariance matrices.

The second research direction relates to how users can be scored as similar or dissimilar. 
An intriguing question raised in the Results and Discussion section is how users can be clustered into groups of similar individuals, where users within each group benefit from transfer learning from one another. This requires identifying a similarity measure between users that can be utilized during BCI calibration to assign a new target user to an appropriate group.
This question also touches upon how transfer learning effectively can be made using multiple source data. Many of the referenced papers in the Related Works section, \cref{sec:related_work}, have used some weighing function or voting classifier to do multi-source transfer learning. However, a transfer learning method similar to the RPA method but based on many similar users would be interesting to find. Another alluring question related to the similarity of users is why we observe that user A is beneficial for user B in a transfer learning setting, while user B is not beneficial for user A, as discussed in a previous subsection of this Conclusions and Future Research section.

\section{Summary}\label{sec:conclusions}
\noindent
During BCI calibration, there is no time to evaluate all source data candidates available for transfer learning. Thus, source data selection is needed. Finding good source data is important, as the final performance of the BCI differs significantly depending on which source data was used during transfer learning. How to select the source data based on features that are available during the BCI calibration is an open research question. 
The results of this paper show that simple features of the target data, available during BCI calibration, can be used to predict the transfer learning performance for a source-target data combination and can thus be effectively applied for source data selection in BCI transfer learning. The presented \myMethod method outperforms other source data selection methods on the studied BCI dataset.

\section*{Contribution}\label{sec:Contribution}
\noindent
\textbf{Frida Heskebeck}: Conceptualization, Methodology, Software, Writing - Original Draft, Writing - Review \& Editing, Visualization, Project administration.\\
\textbf{Carolina Bergeling}: Conceptualization, Methodology, Supervision.\\
\textbf{Bo Bernhardsson}: Conceptualization, Methodology, Writing - Review \& Editing, Supervision, Funding acquisition.

\section*{Acknowledgement}\label{sec:Acknowledgement}
\noindent
Thanks to Pex Tufvesson and Ask Hällström for reading the manuscript and providing feedback on the text.

\printbibliography

@article{MNE_paper,
  title = {{{MEG}} and {{EEG}} Data Analysis with {{MNE}}-{{Python}}},
  author = {Gramfort, Alexandre and Luessi, Martin and Larson, Eric and Engemann, Denis A. and Strohmeier, Daniel and Brodbeck, Christian and Goj, Roman and Jas, Mainak and Brooks, Teon and Parkkonen, Lauri and H{\"a}m{\"a}l{\"a}inen, Matti S.},
  year = {2013},
  volume = {7},
  pages = {1--13},
  doi = {10.3389/fnins.2013.00267},
  journal = {Frontiers in Neuroscience},
  number = {267}
}

@software{MNE_toolbox,
  author       = {Larson, Eric and
                  Gramfort, Alexandre and
                  Engemann, Denis A and
                  Leppakangas, Jaakko and
                  Brodbeck, Christian and
                  Jas, Mainak and
                  Brooks, Teon and
                  Sassenhagen, Jona and
                  Luessi, Martin and
                  McCloy, Daniel and
                  King, Jean-Remi and
                  Höchenberger, Richard and
                  Goj, Roman and
                  Favelier, Guillaume and
                  Brunner, Clemens and
                  van Vliet, Marijn and
                  Wronkiewicz, Mark and
                  Holdgraf, Chris and
                  Rockhill, Alex and
                  Massich, Joan and
                  Bekhti, Yousra and
                  Appelhoff, Stefan and
                  Leggitt, Alan and
                  Dykstra, Andrew and
                  Scheltienne, Mathieu and
                  Luke, Rob and
                  Trachel, Romain and
                  De Santis, Lorenzo and
                  Panda, Asish and
                  Magnuski, Mikołaj and
                  Westner, Britta and
                  Billinger, Martin and
                  Wakeman, Dan G and
                  Strohmeier, Daniel and
                  Bharadwaj, Hari and
                  Linzen, Tal and
                  Barachant, Alexandre and
                  Ruzich, Emily and
                  Bailey, Christopher J and
                  Li, Adam and
                  Moutard, Clément and
                  Bloy, Luke and
                  Raimondo, Fede and
                  Nurminen, Jussi and
                  Montoya, Jair and
                  Woodman, Marmaduke and
                  Lee, Ingoo and
                  Schulz, Martin and
                  Foti, Nick and
                  Nangini, Cathy and
                  García Alanis, José C and
                  Hauk, Olaf and
                  Maddox, Ross and
                  LaPlante, Roan and
                  Drew, Ashley and
                  Dinh, Christoph and
                  Dumas, Guillaume and
                  Huberty, Scott and
                  Hartmann, Thomas and
                  Orfanos, Dimitri Papadopoulos and
                  Ort, Eduard and
                  Benerradi, Johann and
                  Pasler, Paul and
                  Repplinger, Stefan and
                  Rudiuk, Alexander and
                  Radanovic, Ana and
                  Buran, Brad and
                  Massias, Mathurin and
                  Hämäläinen, Matti and
                  Sripad, Praveen and
                  Chirkov, Valerii and
                  Mullins, Christopher and
                  Raimundo, Félix and
                  Alday, Phillip and
                  Pari, Ram and
                  Kornblith, Simon and
                  Halchenko, Yaroslav and
                  Luo, Yu-Han and
                  Kasper, Johannes and
                  Doelling, Keith and
                  Jensen, Mads and
                  Gahlot, Tanay and
                  Nunes, Adonay and
                  Gütlin, Dirk and
                  kjs and
                  Weinstein, Alejandro and
                  Lamus, Camilo and
                  Galván, Catalina María and
                  Moënne-Loccoz, Cristóbal and
                  Heinila, Erkka and
                  Hanna, Jevri and
                  Houck, Jon and
                  Kaneda, Michiru and
                  Klein, Natalie and
                  Kern, Simon and
                  Rantala, Antti and
                  Maess, Burkhard and
                  O'Reilly, Christian and
                  Peterson, Erica and
                  Kolkhorst, Henrich and
                  Banville, Hubert and
                  Maksymenko, Kostiantyn and
                  Clarke, Maggie and
                  Anelli, Matteo and
                  Bannier, Pierre-Antoine and
                  Choudhary, Saket and
                  Gramfort, Alexandre and
                  Forster, Carina and
                  Kim, Cora and
                  Klotzsche, Felix and
                  Wong, Fu-Te and
                  Kojcic, Ivana and
                  Zhang, Jack and
                  Nielsen, Jesper Duemose and
                  Lankinen, Kaisu and
                  Tabavi, Kambiz and
                  Thibault, Louis and
                  Gerster, Moritz and
                  Gayraud, Nathalie and
                  Ward, Nick and
                  Quinn, Andrew and
                  Gauthier, Antoine and
                  Pinsard, Basile and
                  Welke, Dominik and
                  Stephen, Emily and
                  Hornberger, Erik and
                  Hathaway, Evan and
                  Kalenkovich, Evgenii and
                  Mamashli, Fahimeh and
                  Marinato, Giorgio and
                  Anevar, Hafeza and
                  Sosulski, Jan and
                  Stout, Jeff and
                  Calder-Travis, Joshua and
                  Eisenman, Larry and
                  Esch, Lorenz and
                  Dovgialo, Marian and
                  Barascud, Nicolas and
                  Legrand, Nicolas and
                  Falach, Rotem and
                  Deslauriers-Gauthier, Samuel and
                  Cotroneo, Silvia and
                  Matindi, Steve and
                  Bierer, Steven and
                  Férat, Victor and
                  Peterson, Victoria and
                  Baratz, Zvi and
                  Tonin, Alessandro and
                  Kovrig, Alexander and
                  Pascarella, Annalisa and
                  Karekal, Apoorva and
                  de la Torre, Carlos and
                  Gohil, Chetan and
                  Zhao, Christina and
                  Altukhov, Dmitrii and
                  Krzemiński, Dominik and
                  Welke, Dominik and
                  Makowski, Dominique and
                  Mikulan, Ezequiel and
                  Belonosov, Gennadiy and
                  O'Neill, George and
                  Woessner, Jacob and
                  Schiratti, Jean-Baptiste and
                  Evans, Jen and
                  Drew, Jordan and
                  Teves, Joshua and
                  Mathewson, Kyle and
                  Gwilliams, Laura and
                  Varghese, Lenny and
                  Gemein, Lukas and
                  Hecker, Lukas and
                  Lx37 and
                  van Es, Mats and
                  Boggess, Matt and
                  Eberlein, Matthias and
                  Sherif, Mohamed and
                  Kozhemiako, Nataliia and
                  Srinivasan, Naveen and
                  Wilming, Niklas and
                  Chapochnikov, Nikolai and
                  Kozynets, Oleh and
                  Ablin, Pierre and
                  Bertrand, Quentin and
                  Shoorangiz, Reza and
                  Hübner, Rodrigo and
                  Sommariva, Sara and
                  Er, Sena and
                  Khan, Sheraz and
                  Herbst, Sophie and
                  Datta, Sumalyo and
                  Papadopoulo, Theodore and
                  Jochmann, Thomas and
                  Binns, Thomas Samuel and
                  Merk, Timon and
                  Flak, Tod and
                  Dupré la Tour, Tom and
                  Stenner, Tristan and
                  NessAiver, Tziona and
                  akshay0724 and
                  sviter and
                  Earle-Richardson, Aaron and
                  Hindle, Abram and
                  Koutsou, Achilleas and
                  Fecker, Adeline and
                  Wagner, Adina and
                  Ciok, Alex and
                  Pradhan, Aniket and
                  Padee, Anna and
                  Dubarry, Anne-Sophie and
                  Waniek, Anton Nikolas and
                  Singhal, Archit and
                  Rokem, Ariel and
                  Pelzer, Arne and
                  Hurst, Austin and
                  Beasley, Ben and
                  Nicenboim, Bruno and
                  de la Torre, Carlos and
                  Clauss, Christian and
                  Mista, Christian and
                  Li, Chun-Hui and
                  Braboszcz, Claire and
                  Schad, Daniel Carlström and
                  Hasegan, Daniel and
                  Tse, Daniel and
                  Sleiter, Darin Erat and
                  Haslacher, David and
                  Sabbagh, David and
                  Kostas, Demetres and
                  Petkova, Desislava and
                  Issagaliyeva, Dinara and
                  Das, Diptyajit and
                  Wetzel, Dominik and
                  Eich, Eberhard and
                  DuPre, Elizabeth and
                  Lau, Ellen and
                  Olivetti, Emanuele and
                  Varano, Enrico and
                  Altamiranda, Enzo and
                  Brayet, Eric and
                  de Montalivet, Etienne and
                  Goldstein, Evgeny and
                  Zamberlan, Federico and
                  Pop, Florin and
                  Weber, Frederik D and
                  Tan, Gansheng and
                  Brookshire, Geoff and
                  O'Neill, George and
                  Giulio and
                  Maymandi, Hamid and
                  Abdelhedi, Hamza and
                  Sonntag, Hermann and
                  Ye, Hongjiang and
                  Shin, Hyonyoung and
                  Elmas, Hüseyin Orkun and
                  Machairas, Ilias and
                  Kaczmarzyk, Jakub and
                  Zerfowski, Jan and
                  van den Bosch, Jasper J F and
                  Van Der Donckt, Jeroen and
                  van der Meer, Johan and
                  Niediek, Johannes and
                  Veillette, John and
                  Koen, Josh and
                  Bear, Joshua J and
                  Zhu, Judy D and
                  Dammers, Juergen and
                  Galán, Julia Guiomar Niso and
                  Welzel, Julius and
                  Slama, Katarina and
                  Leinweber, Katrin and
                  Grabot, Laetitia and
                  Andersen, Lau Møller and
                  Barbosa, Leonardo S and
                  Hamilton, Liberty and
                  Alfine, Lorenzo and
                  Hejtmánek, Lukáš and
                  Kitzbichler, Manfred and
                  Kumar, Manoj and
                  Kadwani, Manorama and
                  Sutela, Manu and
                  Koculak, Marcin and
                  Henney, Mark Alexander and
                  van Harmelen, Martin and
                  MartinBaBer and
                  Courtemanche, Matt and
                  Tucker, Matt and
                  Visconti di Oleggio Castello, Matteo and
                  Dold, Matthias and
                  Toivonen, Matti and
                  Shader, Maureen and
                  Cespedes, Mauricio and
                  Krause, Michael and
                  Rybář, Milan and
                  He, Mingjian and
                  Daneshzand, Mohammad and
                  Gensollen, Nicolas and
                  Proulx, Nicole and
                  Focke, Niels and
                  Chalas, Nikolas and
                  Shubi, Omer and
                  Sundaram, Padma and
                  Roujansky, Paul and
                  Silva, Pedro and
                  Molfese, Peter J and
                  Das, Proloy and
                  Li, Quanliang and
                  Barthélemy, Quentin and
                  Nadkarni, Rahul and
                  Gatti, Ramiro and
                  Apariciogarcia, Ramonapariciog and
                  Nasri, Reza and
                  Koehler, Richard and
                  Stargardsky, Riessarius and
                  Oostenveld, Robert and
                  Seymour, Robert and
                  Schirrmeister, Robin Tibor and
                  Law, Ryan and
                  Pai, Sagun and
                  Perry, Sam and
                  Louviot, Samuel and
                  Ruuskanen, Santeri and
                  Saha, Sawradip and
                  Mathot, Sebastiaan and
                  Major, Sebastian and
                  Treguer, Sebastien and
                  Castaño, Sebastián and
                  Deng, Senwen and
                  Antopolskiy, Sergey and
                  Wong, Simeon and
                  Wong, Simeon and
                  Poil, Simon-Shlomo and
                  Foslien, Sondre and
                  Singh, Sourav and
                  Chambon, Stanislas and
                  Bethard, Steven and
                  Gutstein, Steven M and
                  Meyer, Svea Marie and
                  Wang, T and
                  Donoghue, Thomas and
                  Moreau, Thomas and
                  Radman, Thomas and
                  Gates, Timothy and
                  Ma, Tom and
                  Stone, Tom and
                  Clausner, Tommy and
                  Anijärv, Toomas Erik and
                  Xia, Xiaokai and
                  Zuo, Yiping and
                  Zhang, Zhi and
                  buildqa and
                  luzpaz},
  title        = {MNE-Python},
  month        = sep,
  year         = 2023,
  publisher    = {Zenodo},
  version      = {v1.5.1},
  doi          = {10.5281/zenodo.8322569},
  url          = {https://mne.tools},
urldate     = {2024-03-15}
}

@software{pyriemann,
  author       = {Alexandre Barachant and
                  Quentin Barthélemy and
                  Jean-Rémi King and
                  Alexandre Gramfort and
                  Sylvain Chevallier and
                  Pedro L. C. Rodrigues and
                  Emanuele Olivetti and
                  Vladislav Goncharenko and
                  Gabriel Wagner vom Berg and
                  Ghiles Reguig and
                  Arthur Lebeurrier and
                  Erik Bjäreholt and
                  Maria Sayu Yamamoto and
                  Pierre Clisson and
                  Marie-Constance Corsi},
  title        = {pyRiemann/pyRiemann: v0.5},
  year         = 2024,
  publisher    = {Zenodo},
  version      = {v0.5},
  doi          = {10.5281/zenodo.8059038},
  url          = {https://pyriemann.readthedocs.io},
  urldate      = {2024-03-15}
}

@incollection{nam_braincomputer_2018-ch1,
  title = {Brain--{{Computer Interface}}: {{An Emerging Interaction Technology}}},
  booktitle = {Brain--{{Computer Interfaces Handbook}}: {{Technological}} and {{Theoretical Advances}}},
  author = {Nam, Chang S and Choi, Imchul and Wadeson, Amy and Whang, Mincheol},
  editor = {Nam, Chang S. and Nijholt, Anton and Lotte, Fabien},
  year = {2018},
  month = jan,
  pages = {12--52},
  publisher = {CRC Press},
  address = {New York},
  doi = {10.1201/9781351231954},
  isbn = {978-1-351-23195-4}
}

@article{schalk_bci2000_2004,
  title = {{{BCI2000}}: A General-Purpose Brain-Computer Interface ({{BCI}}) System},
  shorttitle = {{{BCI2000}}},
  author = {Schalk, Gerwin and McFarland, Dennis J. and Hinterberger, Thilo and Birbaumer, Niels and Wolpaw, Jonathan R.},
  year = {2004},
  month = jun,
  journal = {IEEE Trans Biomed Eng},
  volume = {51},
  number = {6},
  pages = {1034--1043},
  issn = {0018-9294},
  doi = {10.1109/TBME.2004.827072},
  langid = {english},
  pmid = {15188875},
    url = {https://www.physionet.org/content/eegmmidb/1.0.0/},
urldate = {2024-03-15}
}

@inproceedings{coelho_rodrigues_when_2019,
  title = {''{{When}} Does It Work?'': {{An}} Exploratory Analysis of Transfer Learning for {{BCI}}},
  shorttitle = {''{{When}} Does It Work?},
  booktitle = {{{BCI}} 2019 - 8th {{International Brain-Computer Interface Conference}}},
  author = {Coelho Rodrigues, Pedro Luiz and Congedo, Marco and Jutten, Christian},
  year = {2019},
  month = sep,
  address = {Graz, Austria},
  urldate = {2024-04-09}
}

@article{lotte_signal_2015,
  title = {Signal {{Processing Approaches}} to {{Minimize}} or {{Suppress Calibration Time}} in {{Oscillatory Activity-Based Brain}}--{{Computer Interfaces}}},
  author = {Lotte, Fabien},
  year = {2015},
  month = jun,
  journal = {Proceedings of the IEEE},
  volume = {103},
  number = {6},
  pages = {871--890},
  issn = {1558-2256},
  doi = {10.1109/JPROC.2015.2404941}
}

@incollection{nam_transfer_2018,
  title = {Transfer {{Learning}} for {{BCIs}}},
  booktitle = {Brain--{{Computer Interfaces Handbook}}: {{Technological}} and {{Theoretical Advances}}},
  author = {Jayaram, Vinay and Fiebig, Karl-Heinz and Peters, Jan and {Grosse-Wentrup}, Moritz},
  editor = {Nam, Chang S. and Nijholt, Anton and Lotte, Fabien},
  year = {2018},
  month = jan,
  pages = {425--441},
  publisher = {CRC Press},
  address = {New York},
  doi = {10.1201/9781351231954},
  isbn = {978-1-351-23195-4}
}

@article{rodrigues_riemannian_2019,
  title = {Riemannian {{Procrustes Analysis}}: {{Transfer Learning}} for {{Brain}}--{{Computer Interfaces}}},
  shorttitle = {Riemannian {{Procrustes Analysis}}},
  author = {Rodrigues, Pedro Luiz Coelho and Jutten, Christian and Congedo, Marco},
  year = {2019},
  month = aug,
  journal = {IEEE Transactions on Biomedical Engineering},
  volume = {66},
  number = {8},
  pages = {2390--2401},
  issn = {1558-2531},
  doi = {10.1109/TBME.2018.2889705}
}

@article{wan_review_2021,
  title = {A Review on Transfer Learning in {{EEG}} Signal Analysis},
  author = {Wan, Zitong and Yang, Rui and Huang, Mengjie and Zeng, Nianyin and Liu, Xiaohui},
  year = {2021},
  month = jan,
  journal = {Neurocomputing},
  volume = {421},
  pages = {1--14},
  issn = {0925-2312},
  doi = {10.1016/j.neucom.2020.09.017},
  urldate = {2024-01-12}
}

@article{yger_riemannian_2017,
  title = {Riemannian {{Approaches}} in {{Brain-Computer Interfaces}}: {{A Review}}},
  shorttitle = {Riemannian {{Approaches}} in {{Brain-Computer Interfaces}}},
  author = {Yger, Florian and Berar, Maxime and Lotte, Fabien},
  year = {2017},
  month = oct,
  journal = {IEEE Transactions on Neural Systems and Rehabilitation Engineering},
  volume = {25},
  number = {10},
  pages = {1753--1762},
  issn = {1558-0210},
  doi = {10.1109/TNSRE.2016.2627016},
  urldate = {2024-01-12}
}

@phdthesis{heskebeck_calibration_2023,
  type = {Licentiate {{Thesis}}},
  title = {On {{Calibration Algorithms}} for {{Real-Time Brain-Computer Interfaces}}},
  author = {Heskebeck, Frida},
  year = {2023},
  month = oct,
  address = {Lund},
  school = {Department of Automatic Control, Lund Institute of Technology, Lund University}
}

@article{scikit-learn,
  title={Scikit-learn: Machine Learning in {P}ython},
  author={Pedregosa, F. and Varoquaux, G. and Gramfort, A. and Michel, V.
          and Thirion, B. and Grisel, O. and Blondel, M. and Prettenhofer, P.
          and Weiss, R. and Dubourg, V. and Vanderplas, J. and Passos, A. and
          Cournapeau, D. and Brucher, M. and Perrot, M. and Duchesnay, E.},
  journal={Journal of Machine Learning Research},
  volume={12},
  pages={2825--2830},
  year={2011}
}

@inproceedings{jeon_domain_2019,
  title = {Domain {{Adaptation}} with {{Source Selection}} for {{Motor-Imagery}} Based {{BCI}}},
  booktitle = {2019 7th {{International Winter Conference}} on {{Brain-Computer Interface}} ({{BCI}})},
  author = {Jeon, Eunjin and Ko, Wonjun and Suk, Heung-Il},
  year = {2019},
  month = feb,
  pages = {1--4},
  issn = {2572-7672},
  doi = {10.1109/IWW-BCI.2019.8737340},
  urldate = {2024-07-04}
}

@article{li_transfer_2021,
  title = {Transfer {{Learning Based}} on {{Hybrid Riemannian}} and {{Euclidean Space Data Alignment}} and {{Subject Selection}} in {{Brain-Computer Interfaces}}},
  author = {Li, Yan and Wei, Qingguo and Chen, Yuebin and Zhou, Xichen},
  year = {2021},
  journal = {IEEE Access},
  volume = {9},
  pages = {6201--6212},
  issn = {2169-3536},
  doi = {10.1109/ACCESS.2020.3048683},
  urldate = {2024-07-04}
}

@inproceedings{orihara_active_2023,
  title = {Active {{Selection}} of {{Source Patients}} in {{Transfer Learning}} for {{Epileptic Seizure Detection Using Riemannian Manifold}}},
  booktitle = {{{ICASSP}} 2023 - 2023 {{IEEE International Conference}} on {{Acoustics}}, {{Speech}} and {{Signal Processing}} ({{ICASSP}})},
  author = {Orihara, Toshiki and Hassan, Kazi Mahmudul and Tanaka, Toshihisa},
  year = {2023},
  month = jun,
  pages = {1--5},
  issn = {2379-190X},
  doi = {10.1109/ICASSP49357.2023.10095272},
  urldate = {2024-07-04}
}

@article{wei_multi-source_2023,
  title = {A {{Multi-Source Transfer Joint Matching Method}} for {{Inter-Subject Motor Imagery Decoding}}},
  author = {Wei, Fulin and Xu, Xueyuan and Jia, Tianyuan and Zhang, Daoqiang and Wu, Xia},
  year = {2023},
  journal = {IEEE Transactions on Neural Systems and Rehabilitation Engineering},
  volume = {31},
  pages = {1258--1267},
  issn = {1558-0210},
  doi = {10.1109/TNSRE.2023.3243257},
  urldate = {2024-07-04}
}

@article{wu_multi-source_2023,
  title = {Multi-Source Online Transfer Algorithm Based on Source Domain Selection for {{EEG}} Classification},
  author = {Wu, Zizhuo and She, Qingshan and Hou, Zhelong and Li, Zhenyu and Tian, Kun and Ma, Yuliang},
  year = {2023},
  month = jan,
  journal = {Mathematical Biosciences and Engineering},
  volume = {20},
  number = {3},
  pages = {4560--4573},
  publisher = {AIMS Press},
  issn = {1551-0018},
  doi = {10.3934/mbe.2023211},
  urldate = {2024-01-31},
  langid = {english}
}

@article{xu_selective_2021,
  title = {Selective {{Cross-Subject Transfer Learning Based}} on {{Riemannian Tangent Space}} for {{Motor Imagery Brain-Computer Interface}}},
  author = {Xu, Yilu and Huang, Xin and Lan, Quan},
  year = {2021},
  month = nov,
  journal = {Front. Neurosci.},
  volume = {15},
  publisher = {Frontiers},
  issn = {1662-453X},
  doi = {10.3389/fnins.2021.779231},
  urldate = {2024-07-04},
  langid = {english}
}

\end{document}